\documentstyle[aps,preprint]{revtex}

\widetext
\draft
\tighten
\begin{document}

\preprint{EFUAZ FT-95-15}

\title{Lagrangian for the Majorana-Ahluwalia
Construct\thanks{Submitted to ``{\it Nuovo Cim. A}".}}

\author{{\bf Valeri V. Dvoeglazov}\thanks{On leave of absence from
{\it Dept. Theor. \& Nucl. Phys., Saratov State University,
Astrakhanskaya ul., 83, Saratov\, RUSSIA.}\,
Internet address: dvoeglazov@main1.jinr.dubna.su}}

\address {
Escuela de F\'{\i}sica, Universidad Aut\'onoma de Zacatecas \\
Antonio Doval\'{\i} Jaime\, s/n, Zacatecas 98000, ZAC., M\'exico\\
Internet address:  VALERI@CANTERA.REDUAZ.MX
}

\date{May 31, 1995}

\maketitle

\begin{abstract}
The equations describing self/anti-self charge conjugate states,
recently proposed by Ahluwalia, are re-written to covariant form.
The corresponding Lagrangian for the neutral particle theory is
proposed.  From a group-theoretical viewpoint the construct is
an example  of the Nigam-Foldy-Bargmann-Wightman-Wigner-type
quantum field theory based on the doubled representations of the extended
Lorentz group.  Relations with the Sachs-Schwebel and Ziino-Barut concepts
of relativistic quantum theory are discussed.
\end{abstract}

\pacs{PACS numbers: 11.10.Ef, 11.30.Cp}

\newpage

\renewcommand{\thefootnote}{\alph{footnote}}
\baselineskip13pt

Recently, the Majorana-McLennan-Case construct for
neutrino and photon~\cite{Majorana,Case} got a substantial development in
the works of Ahluwalia {\it et al.}~\cite{DVA}. In connection with
observation of candidate events for neutrino $\bar \nu_\mu \rightarrow
\bar \nu_e$ oscillations at LSND LAMPF~\cite{LSND}, that are not predicted
and are not explained by gauge theories of  the Glashow-Weinberg-Salam
(GWS) type~\cite{WSG}, an alternative insight in neutral particle physics
has some reasons.  While thirty years passed since the proposal of the GWS
model, we are still far from understanding many its essential
theoretical ingredients; first of all, the fundamental origins of
``parity violation" effect~\cite{PV}, the Kobayashi-Maskawa
mixing~\cite{KM} and Higgs phenomenon~\cite{Higgs}.  Experimental neutrino
physics and astrophysics provided us by new puzzles~\cite{NR}, that until
now did not find adequate explanation.

In the mean time, the Majorana-McLennan-Case-Ahluwalia construct is based
on a very natural principle for describing neutral particles: a principle
of self/anti-self charge conjugacy of the states that correspond to
neutrinos, $j=1$ and higher spin neutral particles. The kinematical
framework of the theory has been given in ref.~\cite{DVA}.  Unusual
properties of the construct, such as ``incompatibility of simultaneous
existence of self/anti-self charge conjugacy and helicity eigenstates",
impossibility of the ``standard fashion" gauge interaction,
bi-orthonormality of physical states and the remarkable $\lambda^{S,A}
\leftrightarrow \rho^{A,S}$ property with respect to space reflection,
have been revealed.  The 4-spinors used there (that describe
self/anti-self charge conjugate states) are the following:
\begin{equation}
\lambda(p^\mu)\,\equiv \pmatrix{ \left (
\zeta_\lambda\,\Theta_{[j]}\right )\,\phi^\ast_{_L}(p^\mu)\cr
\phi_{_L}(p^\mu)} \,\,,\quad \rho(p^\mu)\,\equiv \pmatrix{
\phi_{_R}(p^\mu)\cr
\left ( \zeta_\rho\,\Theta_{[j]}\right )^\ast
\,\phi^\ast_{_R}(p^\mu)} \,\,\quad .\label{sp-dva}
\end{equation}
$\zeta_\lambda$ and $\zeta_\rho$ are the phase factors that
are  fixed by the conditions of  self/anti-self conjugacy,
$\Theta_{[j]}$ is the Wigner time-reversal operator for spin $j$.
They are called usually {\it the Majorana-(like) spinors}.

Next, in the papers~\cite{DVA} the following equation for
$\lambda^{S,A} (p^\mu)$ has been presented:
\begin{eqnarray}
\pmatrix{-\,\openone & \zeta_\lambda\,\exp\left(
{\bf J}\,\cdot \bbox{\varphi}\right )
\,\Theta_{[j]}\,{\mit\Xi}_{[j]}\, \exp\left( {\bf J}\,\cdot \bbox{\varphi}
\right )\cr
\zeta_\lambda\,\exp\left(-\, {\bf J}\,\cdot\bbox{\varphi}\right)
\,{\mit\Xi}^{-1}_{[j]}\,\Theta_{[j]}\,
\exp\left(- \,{\bf J}\,\cdot\bbox{\varphi}
\right) & -\,\openone}\,\lambda (p^\mu)\,=\,0.\label{genweq1}
\end{eqnarray}
The analogous equation for $\rho^{S,A} (p^\mu)$  is:
\begin{eqnarray}
\pmatrix{-\,\openone & \zeta_\rho^*\,\exp\left(
{\bf J}\,\cdot \bbox{\varphi}\right )\,
{\mit\Xi}_{[j]}^{-1}\,\Theta_{[j]}\, \exp\left( {\bf J}\,\cdot
\bbox{\varphi} \right )\cr
\zeta_\rho^*\,\exp\left(-\, {\bf
J}\,\cdot\bbox{\varphi}\right) \,\Theta_{[j]}\,{\mit\Xi}_{[j]}\,
\exp\left(- \,{\bf J}\,\cdot\bbox{\varphi}
\right) & -\,\openone}\,\rho (p^\mu)\,=\,0,\label{genweq2}
\end{eqnarray}
provided that the overall phase factors of $\phi_{_R}^h
(\overcirc{p}^\mu)$
are chosen to be equal to the ones of $\phi_{_L}^h (\overcirc{p}^\mu)$.
${\mit\Xi}_{[j]}$ is the matrix connecting at-rest 2-spinors with
their complex conjugates:
\begin{equation}
\left [\phi_{_{L,R}}^h (\overcirc{p}^\mu)\right ]^* =
{\mit\Xi}_{[j]} \phi_{_{L,R}}^h (\overcirc{p}^\mu)\quad;
\end{equation}
${\bf J}$ are the spin-$j$ matrices; $\overcirc{p}^\mu$ denotes
the at-rest 4-momentum, $\bbox{\varphi}$ are the Lorentz
boost parameters.\footnote{In the paper we try to keep the notation of
ref.~\cite{DVA}. Some important results of the cited paper will be used
below without reference.}

The philosophy of a Lagrangian field theory is
based on realization of the Principle of Least Action, that is assumed
to be valid for all physical systems. ``The dynamics of the system is
specified once the Lagrangian is given"~\cite{Barut0}.
Unfortunately, in ref.~\cite{DVA} field dynamics has not been
presented in details.  The main goal of this Letter is to re-write the
``instant-form" equations for the 4-spinors
$\lambda^{S,A} (p^\mu)$ and $\rho^{S,A} (p^\mu)$, Eqs.
(\ref{genweq1}) and (\ref{genweq2}), to covariant forms and to deduce the
corresponding Lagrangian with the aim to construct the dynamics of ``truly"
neutral particles in the $(1/2,0)\oplus (0,1/2)$ representation space.

First of all, we note that
\begin{eqnarray}
\lefteqn{\Theta_{[1/2]}\, \Xi_{[1/2]} = \Xi^{-1}_{[1/2]} \,\Theta_{[1/2]}\,
=\, i\, \frac{\sigma_1 p_2  - \sigma_2 p_1}{\sqrt{(p + p_3)(p - p_3) }}\,
=\nonumber}\\
&=& \, U_{+} (p^\mu) U_{-} (p^\mu) = -U_+^{-1} (p^\mu) U_-^{-1} (p^\mu)
= - U^{-1}_{\pm} (p^\mu) U_{\pm}
(\tilde p^{\mu}) = U^{-1}_{\pm} (\tilde p^{\mu}) U_{\pm} (p^\mu)\quad,
\end{eqnarray}
where $U_{\pm} (p^\mu)$ are the matrices of the $2\times 2$ unitary
transformation  to the helicity
representation~[11a,Eq.(3.1)] and ref.~\cite[p.71]{Novozh}:
\begin{equation}
U_+ (p^\mu) \sigma_3 U^{-1}_+ (p^\mu) \,=\, \frac{({\bbox\sigma}\cdot {\bf
p})}{p}\quad, \quad
U_-^{-1} (p^\mu) \sigma_3 U_- (p^\mu) \,= \,
-\,\frac{({\bbox\sigma}\cdot {\bf p})}{p}\quad;\label{hr1}
\end{equation}
$p=\vert {\bf p}\vert =\sqrt{E^2 -m ^2}$ and $\tilde p^{\mu}$ is the
parity-conjugated momentum.  Using Eq. (\ref{hr1}) one can obtain
the following equations for 4-spinors of the second kind
$\lambda_{_{H}} (p^\mu) = {\cal U} \lambda
(p^\mu)$ and $\rho_{_{H}} (p^\mu) ={\cal U} \rho (p^\mu)$,
$\tilde \lambda_{_{H}} (p^\mu) = \tilde {\cal U} \lambda
(p^\mu)$ and $\tilde \rho_{_{H}} (p^\mu) = \tilde {\cal U} \rho (p^\mu)$
in the new representations:\footnote{We have used that
$$\exp ({\pm{\bbox{\sigma}\over 2}\cdot
\bbox{\varphi}}) = \cosh {\varphi \over 2} \pm (\bbox{\sigma}\cdot
\bbox{\hat \varphi}) \sinh {\varphi \over 2}\quad.$$
The equations for $\rho_H (p^\mu)$ and $\tilde \rho_H (p^\mu)$ are obtained
from (7a,b) after the substitutions
$\zeta_\lambda \rightarrow \zeta^\ast_\rho$.}
\begin{mathletters}
\begin{eqnarray}
\left
[\zeta_\lambda (\cosh^2 {\varphi\over 2} -\sinh^2 {\varphi \over 2})
\gamma_5 \gamma_0 - \openone \right ]
\lambda_{_{H}} (p^\mu) \, &=& \,0\quad,\label{equ1}\\
\left
[\zeta_\lambda (\cosh^2 {\varphi\over 2} -\sinh^2 {\varphi \over 2})
\gamma_5 \gamma_0 + \openone \right ]
\tilde \lambda_{_{H}} (p^\mu) \, &=& \,0\quad.\label{equ2}
\end{eqnarray}
\end{mathletters}
The unitary matrices ${\cal U}$ are implied to be of the
following set:
\begin{eqnarray}
{\cal U}_\pm = \pmatrix{ U_\pm (\tilde
p^{\mu}) & 0\cr 0 & U_\pm (p^\mu)\cr}\quad,\quad
\widetilde {\cal U}_\pm =
\pmatrix{ U_\pm (p^{\mu}) & 0\cr 0 & U_\pm (\tilde
p^\mu)\cr}\quad.\quad\label{Eq}
\end{eqnarray}
The transformation to  the helicity representation is a rotation, indeed,
that, according to the ordinary viewpoint, can not have influence
physical results. Let me mention that the equations similar to
Eqs. (\ref{equ1},\ref{equ2}) could also be obtained after
calculations  with the transformation matrix $\Omega ({1\over 2})$,
or $\Omega^T ({1\over 2})$, the matrix of a transfer to the light-front
representation.  For instance, one can re-write Eq. (\ref{genweq1})
to the following {\it non-dynamical} form ( $h$ is the helicity;
$(\bbox{\sigma}\cdot \widehat {\bf p}) \phi_{_{L,R}} (p^\mu) = h
\phi_{_{L,R}}$):
\begin{eqnarray}
\pmatrix{-\openone & - \zeta_\lambda (h \sigma_3 p -
p_3)/\sqrt{p_r p_l}\cr + \zeta_\lambda (h \sigma_3 p + p_3)/\sqrt{p_r p_l}
& -\openone\cr} \lambda (p^\mu) = 0\quad.
\end{eqnarray}
It is not surprising since the famous Melosh transformation,
ref.~\cite{Melosh,DVALF} (see also~\cite{Novo}),
\begin{eqnarray}
\Omega ({1\over2}) = \frac{1}{\left [ 2 (E+m) p^+ \right ]^{1/2}}
\pmatrix{\beta ({1\over 2}) & 0\cr 0 & \beta ({1\over 2})\cr}\quad,
\quad \beta ({1\over 2}) =
\pmatrix{p^+ + m & - p_r\cr p_l & p^+ + m\cr}\quad,\label{lf}
\end{eqnarray}
is shown  in ref.~\cite{KT} to be a rotation too.\footnote{Let me note
that definitions of Melosh and Ahluwalia for spin-$1/2$ are connected in
the following way: $\,S_{\ _{ref.~\cite{Melosh}}} \sim \beta^T_{\
_{ref.~\cite{DVALF}}}$ \, within an accuracy of normalization and
with the Pauli matrices $\bbox{\sigma}$ are in the standard
representation. The definitions of Kondratyuk and Terent'ev, as follows:
$U_{\ _{ref.~\cite{KT}}} (p) \sim \beta^\ast_{\ _{ref.~\cite{DVALF}}}$.
$T$ stands for transpose operation, the asterisk, for complex
conjugation.} As a matter of fact, these results hint that neutral
particle states can ``live" on the light cones only.

Nevertheless, one can still deduce dynamical
equations for $\lambda^{S,A} (p^\mu)$.
{}From the analysis of the rest spinors, Eqs. (22a,b) of
ref.~[3d] one can find another form of the Ryder-Burgard
relation\footnote{Different generalizations of the
Ryder-Burgard relation for left- and right- at-rest
spinors $\phi_R (\overcirc{p}^\mu)=\pm \phi_L
(\overcirc{p}^\mu)$, so called by Ahluwalia, have also been discussed in
refs.~\cite{Faustov,DVO952,DVO953}.} for the $j=1/2$ case:
\begin{equation}\label{rbug12}
\left [\phi_{_L}^h (\overcirc{p}^\mu)\right ]^* = (-1)^{1/2-h}\,
e^{-i(\theta_1 +\theta_2)}
\,\Theta_{[1/2]} \,\phi_{_L}^{-h} (\overcirc{p}^\mu)\quad.
\end{equation}
Provided that the overall phase factors of at-rest spinors are chosen
to be $\theta_1 +\theta_2 = 0$ we come to the equations:
\begin{mathletters}
\begin{eqnarray}\label{eqq}
\left [{i\over m}\gamma_5\hat p - 1 \right ]
\Upsilon_\pm (p^\mu) &=& 0\quad,\quad\\
\left [{i\over m}\gamma_5\hat p + 1 \right ]
{\cal B}_\pm (p^\mu) &=& 0\quad.
\end{eqnarray}
\end{mathletters}
Here we defined  4-spinors, that are in helicity
eigenstates~\cite{DVO951,DVO953}:
\begin{eqnarray}
\Upsilon_\pm (p^\mu) = \pmatrix{\pm i\Theta_{1/2}
\left [\phi_{_L}^{\mp 1/2} (p^\mu)\right ]^*\cr
\phi_{_L}^{\pm 1/2} (p^\mu)\cr}\quad,\quad
{\cal B}_\pm (p^\mu) = \pmatrix{\mp i\Theta_{1/2}
\left [\phi_{_L}^{\mp 1/2} (p^\mu)\right ]^*\cr
\phi_{_L}^{\pm 1/2} (p^\mu)\cr}\quad.
\end{eqnarray}
Of course, we could start from the equation (\ref{genweq2}) and
obtain the equivalent set:
\begin{eqnarray}
\widetilde \Upsilon_\pm (p^\mu)=
\pmatrix{\phi_{_R}^{\pm 1/2} (p^\mu)\cr
\mp i\Theta_{1/2} \left [\phi_{_R}^{\mp 1/2} (p^\mu)\right
]^*\cr}\quad,\quad
\widetilde {\cal B}_\pm (p^\mu) = \pmatrix{\phi_{_R}^{\pm
1/2}(p^\mu)\cr \pm i\Theta_{1/2} \left [\phi_{_R}^{\mp 1/2} (p^\mu)\right
]^*\cr}\quad.
\end{eqnarray}
The latter can differ from the former only by a phase factor
$e^{if_{_{\pm}}}$ provided
that we keep the ordinary normalization of the Pauli $\phi_{_{L,R}}$
spinors.\footnote{In ref.~\cite{DVO953} we have used the slightly
different notation:  $\Upsilon_\pm (p^\mu) \equiv \pm {\cal B}^{(2)}_\pm
(p^\mu)$, \, ${\cal B}_\pm (p^\mu) \equiv \pm \Upsilon_\pm^{(2)} (p^\mu)$;
and
$\widetilde \Upsilon_\pm (p^\mu) \equiv \pm \Upsilon^{(1)}_\pm (p^\mu)$, \,
$\widetilde {\cal B}_\pm (p^\mu) \equiv \pm {\cal B}_\pm^{(1)} (p^\mu)$.}

One can then use the relations between the 4-spinors $\Upsilon (p^\mu)$,
\,${\cal B} (p^\mu)$ and $\lambda (p^\mu)$:\footnote{The arrows $\uparrow
\downarrow$ should be referred to `chiral helicity' introduced
in ref.~\cite{DVA}.}
\begin{mathletters}
\begin{eqnarray}
\Upsilon_+ (p^\mu) &=& \pm\frac{1+\gamma_5}{2} \lambda^{S,A}_\downarrow
+ \frac{1-\gamma_5}{2} \lambda^{S,A}_\uparrow\quad,\\
\Upsilon_- (p^\mu) &=& \mp\frac{1+\gamma_5}{2} \lambda^{S,A}_\uparrow
+ \frac{1-\gamma_5}{2} \lambda^{S,A}_\downarrow\quad,\\
{\cal B}_+ (p^\mu) &=& \mp\frac{1+\gamma_5}{2} \lambda^{S,A}_\downarrow
+ \frac{1-\gamma_5}{2} \lambda^{S,A}_\uparrow\quad,\\
{\cal B}_- (p^\mu) &=& \pm\frac{1+\gamma_5}{2} \lambda^{S,A}_\uparrow
+ \frac{1-\gamma_5}{2} \lambda^{S,A}_\downarrow\quad;
\end{eqnarray}
\end{mathletters}
and between the 4-spinors $\widetilde \Upsilon (p^\mu)$,
$\widetilde {\cal B} (p^\mu)$ and $\rho (p^\mu)$:
\begin{mathletters}
\begin{eqnarray}
\widetilde\Upsilon_+ (p^\mu) &=& \left [\frac{1+\gamma_5}{2}
\rho^{S,A}_\uparrow \pm \frac{1-\gamma_5}{2}
\rho^{S,A}_\downarrow\right ]\quad,\\
\widetilde\Upsilon_- (p^\mu) &=& \left [
\frac{1+\gamma_5}{2} \rho^{S,A}_\downarrow \mp \frac{1-\gamma_5}{2}
\rho^{S,A}_\uparrow \right ]\quad,\\
\widetilde {\cal B}_+ (p^\mu) &=& \left [
\frac{1+\gamma_5}{2} \rho^{S,A}_\uparrow \mp \frac{1-\gamma_5}{2}
\rho^{S,A}_\downarrow\right ]\quad,\\
\widetilde {\cal B}_- (p^\mu) &=& \left [
\frac{1+\gamma_5}{2} \rho^{S,A}_\downarrow \pm \frac{1-\gamma_5}{2}
\rho^{S,A}_\uparrow\right ]\quad.
\end{eqnarray}
\end{mathletters}
Furthermore, we assume that the parity violation is {\it not} explicit
in the meaning of ref.~\cite{Ziino} and one can use Eqs. (40a,b) of
ref.~[3c]. Finally, if imply like ref.~[3d] that the $\lambda^S (p^\mu)$
(and, therefore,  $\rho^A (p^\mu)$, see Eqs. (\ref{genweq1},\ref{genweq2}))
are the solutions corresponding to positive frequencies; and
$\lambda^A (p^\mu)$ and $\rho^S (p^\mu)$, to
negative frequencies, the equations for 4-spinors of the
same `chiral helicity' take the following ``mad" forms
comparing with the ordinary Dirac case~\cite{Dirac1}:\footnote{Of course,
one can re-write the obtained equations to two-component form
for $\phi_R (p^\mu)$, $\phi_L (p^\mu)$, $i\Theta_{[1/2]} \phi_R^\ast
(p^\mu)$ and $i\Theta_{[1/2]} \phi_L^\ast (p^\mu)$. {\it Cf.} with the
equations (11a,b) of ref.~[2b].}
\begin{mathletters}
\begin{eqnarray}
i \gamma^\mu \partial_\mu \lambda^S (x) - m \rho^A (x) &=& 0 \quad,
\label{11}\\
i \gamma^\mu \partial_\mu \rho^A (x) - m \lambda^S (x) &=& 0 \quad;
\label{12}
\end{eqnarray}
\end{mathletters}
and
\begin{mathletters}
\begin{eqnarray}
i \gamma^\mu \partial_\mu \lambda^A (x) + m \rho^S (x) &=& 0\quad,\\
\label{13}
i \gamma^\mu \partial_\mu \rho^S (x) + m \lambda^A (x) &=& 0\quad.
\label{14}
\end{eqnarray}
\end{mathletters}
They can be written in the 8-component form as follows:
\begin{mathletters}
\begin{eqnarray}
\left [i \Gamma^\mu \partial_\mu - m\right ] \Psi_{(+)} (x) &=& 0\quad,
\label{psi1}\\
\left [i \Gamma^\mu \partial_\mu + m\right ] \Psi_{(-)} (x) &=& 0\quad,
\label{psi2}
\end{eqnarray}
\end{mathletters}
where we defined the Weinberg {\it dibispinors}\,:\footnote{This name
has been used in my previous works
for the set of $F_{\mu\nu}$ and its dual $\tilde F_{\mu\nu}$,
the wave functions (operators) of the antisymmetric tensor field.
I take a liberty to apply it for the 8-component wave functions
of the $(1/2,0)\oplus (0,1/2)$ representation space too, taking into
account the significant contribution of Dr. Weinberg to modern
physics in his pioneer works of 1964-69, ref.~\cite{Weinberg1}.}
\begin{eqnarray}
\Psi_{(+)} (x) = \pmatrix{\rho^A (x)\cr
\lambda^S (x)\cr}\quad,\quad
\Psi_{(-)} (x) = \pmatrix{\rho^S (x)\cr
\lambda^A (x)\cr}\quad.
\end{eqnarray}
The set of $8\times 8$- component $\Gamma$-
and $T$-matrices\footnote{Similar set of matrices has been defined in
ref.~[24,25c].} is written as
\begin{eqnarray}
\Gamma^\mu =\pmatrix{0 & \gamma^\mu\cr
\gamma^\mu & 0\cr}\quad,\quad \Gamma^5 = \pmatrix{\gamma^5 & 0\cr
0 &\gamma^5\cr}\quad,\quad {\L}^5 = \pmatrix{\gamma^5 & 0\cr
0 & -\gamma^5\cr}\quad,
\end{eqnarray}
\begin{eqnarray}
T_{11} = \pm i\,\pmatrix{\openone & 0 \cr
0 & -\openone\cr}\quad,\quad
T_{01} = \pmatrix{0& \openone \cr
\openone & 0\cr}\quad,\quad
T_{10} = \pm i \, \pmatrix{0 & \openone\cr
- \openone & 0\cr}\quad,\quad
\end{eqnarray}
The latter are defined within an accuracy of the
factors $(-1)^k$. The set is analogous to the Pauli sets of matrices.
$\gamma^\mu$ are the ordinary Dirac matrices.  The set of
(anti)commutation relations of $\Gamma^\mu$ and $T$ matrices could be
deduced from the well-known relations of the Dirac matrices.  {\it E.
g.}, the important commutation relation is
\begin{equation} \L^5
\Gamma^\nu - \Gamma^\nu \L^5 = 0 \quad.\label{crel}
\end{equation}

Next, one can propose, {\it e.g.}, the following Lagrangian
($\overline \Psi_{(\pm)} \equiv \Psi^\dagger_{(\pm)}
\Gamma_0$):\footnote{This form of the Lagrangian for neutral particles
supports the remark made after Eq. (\ref{lf}). See also the remark
after Eq. (15) in ref.~[2a].}
\begin{eqnarray} {\cal L}^{(1)}
&=& {i\over 2} \left [\,\, \overline\Psi_{(+)}\Gamma^\mu \partial_\mu
\Psi_{(+)} - \partial_\mu \overline\Psi_{(+)}\Gamma^\mu \Psi_{(+)} +
\overline\Psi_{(-)}\Gamma^\mu \partial_\mu \Psi_{(-)} -
\partial_\mu \overline\Psi_{(-)}\Gamma^\mu \Psi_{(-)}\,\,\right ]
-\nonumber\\
&\qquad& - m \left ( \overline
\Psi_{(+)} \Psi_{(+)} - \overline \Psi_{(-)} \Psi_{(-)}  \right
)\quad.\label{Lagr}
\end{eqnarray}
It is useful to note that one can
introduce the following gradient transformations  of the first kind for
$\lambda^{S,A} (x)$ and $\rho^{S,A} (x)$ spinors:\footnote{In
general, phase factors in the gradient transformations of
4-spinors $\lambda$ and $\rho$ could be different. (see forthcoming
papers).}
\begin{mathletters}
\begin{eqnarray}
\lambda^\prime (x)
\rightarrow (\cos \alpha -i\gamma^5 \sin\alpha) \lambda
(x)\quad,\label{g10}\\
\overline \lambda^{\,\prime} (x) \rightarrow
\overline \lambda (x) (\cos \alpha - i\gamma^5
\sin\alpha)\quad,\label{g20}\\
\rho^\prime (x) \rightarrow  (\cos \alpha +
i\gamma^5 \sin\alpha) \rho (x) \quad,\label{g30}\\
\overline \rho^{\,\prime} (x) \rightarrow  \overline \rho (x)
(\cos \alpha + i\gamma^5 \sin\alpha)\quad.\label{g40}
\end{eqnarray}
\end{mathletters}
In terms of the field functions $\Psi_{(\pm)} (x)$ they are written
\begin{mathletters}
\begin{eqnarray}
\Psi^{\,\prime}_{(\pm)} (x) \rightarrow \left ( \cos \alpha +
i \L^5 \sin\alpha \right )
\Psi_{(\pm)} (x)\quad,\label{g1}\\
\overline\Psi_{(\pm)}^{\,\prime} (x)
\rightarrow \overline \Psi_{(\pm)} (x)
\left ( \cos \alpha - i \L^5 \sin\alpha \right )\quad.\label{g2}
\end{eqnarray}
\end{mathletters}
Due to the commutation relation
(\ref{crel}) the Lagrangian (\ref{Lagr})  is invariant with respect
to (\ref{g1}-\ref{g2}). Using the analogy with ordinary
quantum electrodynamics the local gradient transformations
(gauge transformations) could also be defined.
Like the ordinary case we are forced to introduce the compensating
field of the vector potential, but in the case under consideration
the covariant derivative is introduced in a slightly different
fashion:
\begin{eqnarray}
&& \partial_\mu \rightarrow \nabla_\mu = \partial_\mu - ig \L^5
A_\mu\quad,\\
&& A_\mu^\prime (x) \rightarrow A_\mu + {1\over g} \,\partial_\mu \alpha
\quad.
\end{eqnarray}
This tells us that self/anti-self conjugate states (more precisely,
the Weinberg dibispinor) can possess the axial charge.

We would like to point out connection of the construct proposed in
ref.~\cite{DVA} with the similar formulations met in the
literature~\cite{Markov,Brana}, see also~\cite{PB}, and, in particular,
with the formulation of refs.~\cite{Barut1,Ziino}.
The asymptotically ``chiral" massive fermions
proposed by Dr. Ziino, are essentially the fields $\lambda^A \sim
\psi_{f}^{ch}$ and $\rho^S \sim \psi_{\bar f}^{ch}$.
It is possible to find the corresponding relations between $\lambda^S,
\rho^A$ of ref.~\cite{DVA} and Ziino's fields of the opposite frequencies.
Barut and Ziino noted on the needed
modifications of our understanding ~[21c] the concept of the
quantization space: ``Such a Fock space should have the manifestly
covariant structure \begin{equation} {\cal F} \equiv {\cal F}^0 \otimes
S_{in}\quad, \end{equation} where ${\cal F}^0$ is an ordinary Fock space
for one {\it indistinct} type of positive- and negative-energy identical
spin-${1\over 2}$ particles (without regard to the proper-mass sign) and
$S_{in}$ is a two-dimensional internal space spanned by the proper-mass
eigenstates $\vert +m >$, $\vert -m >$ thus {\it doubling} ${\cal F}^0$.
This allows [the Dirac-like fields] to be {\it mixed} if a rotation is
performed in $S_{in}$." As a matter of fact, this phenomenon called
doubling has been discovered  by Wigner~\cite{Wigner}, who enumerated the
irreducible projective representations of the full Poincar\`e group
(including reflections). The ordinary Dirac construct ~\cite{Dirac1}
reflects only a simplest case of the general theory.\footnote{The faults
of the standard approach (first of all, to the interaction problem) seemed
to be realized by Dr.  Dirac himself~\cite{Dirac2}. See also~\cite{Nigam}
for an explicit construct of the charge conjugation representation
for Dirac fields. The example of the FNBWW-type quantum field theory
in the $(1,0)\oplus (0,1)$ has been given in~\cite{DVA00}.}

Then, curiously, an attempt to set up {\it the Majorana-like anzatzen}
in the form:\footnote{The physical content still depends
on the choice of the phase factor $\vartheta$.}
\begin{equation}
\phi_L = e^{i\vartheta}\Theta_{[1/2]}
\phi_R^*\quad,
\end{equation}
where $\vartheta = 0, \pm {\pi \over 2}, \pi$\ ,
on the resulting equations (\ref{11}-\ref{14}) leads to tachyonic
solutions.

Next, if we start to built the neutral particle theory from the 4-spinors
of the first kind we would obtain the Case equations~\cite{Case}. They
coincide with a local limit of the Sachs equation~\cite{Sachs0,Sachs1}:
\begin{equation}
q_\mu \partial_\mu\phi_\alpha - \lambda \Theta_{[1/2]} \phi_\alpha^\ast
=0\quad,\label{Sachs10}
\end{equation}
and its conjugate; the latter follow from his treatment of
quantum theory on the ground of the Einstein's
interpretation, which is an
alternative to the Bohr-Heisenberg viewpoint.
$q_\mu$ is the quaternion, $\lambda$ is a complex parameter
with the dimension of mass.  According to the Sachs
viewpoint ``the field equation (\ref{Sachs10}) represents the inertial mass
appearing as a continuous function, $m=\lambda \hbar/c$, rather than a
constant parameter".  As mentioned in ref.~[34c] such an
interpretation predicts an indefinite spectrum of neutrino masses. From
the other hand, in the recent preprint Prof.  Bilenky with
collaborators~\cite{Bilenky} indicated that ``if the LSND signal is
confirmed, it would mean that\ldots there is no natural hierarchy of
coupling among generations in the lepton sector\ldots If future
experiments confirm the existence of [the atmospheric] neutrino anomaly
and the result of the LSND experiment also\ldots it would be necessary to
assume the existence of an additional \ldots neutrino state besides the
three  active flavor neutrino states". Bilenky's statement provides some
phenomenological grounds to the Sachs theoretical construct.

Next, let us mention that for the first time the analysis of the dynamics,
that follow from the ``doubled" representation of the extended Lorentz
group has been undertaken in ref.~[25c]. The Markov's
set of equations for fermion-antifermion 8-component wave function
turns out to be invariant with respect to the transformation with
$\Gamma^5 T_{01}$
matrix, provided that fermion and its antifermion masses are assumed
equal.  ``Only such kinds of interactions which are not invariant with
respect to these transformations can remove the degeneracy over the bare
particle masses." As a matter of fact, in that preprint Prof. Markov
proposed such a type of interaction that can solve the hierarchy
problem~\cite{Barut2}.\footnote{Recently, a mathematical treatment of the
another model with fields transforming in accordance with the double
representation of the extended Lorentz group has been
undertaken~\cite{DVO953}. It leads to very interesting physical
consequences, such as: in the framework one can obtain both charged and
neutral particles; the formalism admits both commutation and
anticommutation relations for describing one or another states,
there is the ``puzzled" state with zero energy-momentum,
zero charge and the Pauli-Lyuban'sky operator.} Finally, the reader can
wish to reveal transparent connections with the physical content
based on the fundamental principle of indistiguishability of
identical particles, discussed in the paper~\cite[p.195]{Kaplan}.

The main result of this Letter is the Lagrangian for the
Majorana-McLennan-Case-Ahluwalia construct. The wave functions
(field operators) present themselves 8-component dibispinors.
{}From a group-theoretical viewpoint the construct is an example of the
Nigam-Foldy-Bargmann-Wightman-Wigner-type theory based on the doubled
representations of the extended Lorentz  group~\cite{Nigam,Wigner}.
In the approaching papers we are going to find dynamical invariants
following from the proposed Lagrangian, propagators and to built the
Feynman diagram technique for this type of Poincar\`e invariant theories.
Probably, the quantum-field particle dynamics should be constructed on the
base of the kinematical postulates of Faustov, Ryder, Burgard and
Ahluwalia, and with taking into account the ideas worked out by Sachs,
Schwebel, Markov, Ziino, Barut and Pashkov~\cite{Pashkov}.

\acknowledgments
This paper is a continuation of considerable efforts undertaken by
Profs. D. V. Ahluwalia and A. F. Pashkov ``to prove well-founded".
Especially, I thank Prof. A. F. Pashkov for drawing my attention to
refs.~\cite{Berg,Sachs0,Sachs1} and Prof. D. V. Ahluwalia, to
ref.~\cite{DVALF}.

I am grateful to Zacatecas University for professorship.

\end{document}